\documentclass[sigconf,10pt]{acmart}

\setcopyright{rightsretained}

\acmDOI{10.475/123_4}

\acmISBN{978-1-4503-7020-2}

\usepackage{microtype}
\usepackage[htt]{hyphenat}
\usepackage{multirow}
\usepackage{color}

\usepackage{booktabs}       
\usepackage{amsfonts}       
\usepackage{nicefrac}       
\usepackage{microtype}      
\usepackage{xcolor}         

\usepackage{graphicx}
\usepackage{caption}
\usepackage{subcaption}
\usepackage{mwe}

\usepackage{wrapfig}
\usepackage{comment}
\newenvironment{packeditemize}{\begin{list}{$\bullet$}{\setlength{\itemsep}{0.5pt}\addtolength{\labelwidth}{-1pt}\setlength{\leftmargin}{\labelwidth}\setlength{\listparindent}{\parindent}\setlength{\parsep}{1pt}\setlength{\topsep}{0pt}}}{\end{list}}

\copyrightyear{2022}
\acmYear{2022}
\setcopyright{acmcopyright}\acmConference[HotNets '22]{The 21st ACM Workshop on Hot Topics in Networks}{November 14--15, 2022}{Austin, TX, USA}
\acmBooktitle{The 21st ACM Workshop on Hot Topics in Networks (HotNets '22), November 14--15, 2022, Austin, TX, USA}
\acmPrice{15.00}
\acmDOI{10.1145/3563766.3564109}
\acmISBN{978-1-4503-9899-2/22/11}

\begin{document}

\title[Rethinking Data-driven Networking
with Foundation Models]{Rethinking Data-driven Networking
with Foundation Models: Challenges and Opportunities}

\author{Franck Le}
\affiliation{%
  \institution{IBM Research}
}

\author{Mudhakar Srivatsa}
\affiliation{%
  \institution{IBM Research}
}

\author{
Raghu Ganti}
\affiliation{%
  \institution{IBM Research}
}

\author{ Vyas Sekar}
\affiliation{%
  \institution{Carnegie Mellon University}
}

\begin{abstract}

Foundational models have caused a paradigm shift in the way artificial intelligence (AI) systems are built. They have had a major impact in natural language processing (NLP), and several other domains, not only reducing the amount of required labeled data or even eliminating the need for it, but also significantly improving performance on a wide range of tasks.  We argue foundation models  can have a similar profound impact on network traffic analysis, and management. More specifically, we show that network data shares several of the properties that are behind the success of foundational models in linguistics. For example, network data contains rich semantic content, and several of the networking tasks (e.g., traffic classification, generation of protocol implementations from specification text, anomaly detection) can find similar counterparts in NLP (e.g., sentiment analysis, translation from natural language to code, out-of-distribution). However, network settings also present  unique characteristics and challenges that must be overcome. 
Our contribution  is in    highlighting   the  opportunities and challenges at the intersection of foundation models and  networking.

\end{abstract}
%
%



\maketitle
\section{Introduction}


\textit{Foundational models}~\cite{bommasani2021opportunities}  describe the class of models that are behind the current paradigm shift for building artificial intelligence (AI) systems where neural networks that are trained on large corpora of unlabeled data, and then adapted to a wide range of downstream tasks with minimal fine-tuning. For example, the BERT (Bidirectional Encoder Representations from
Transformers) model~\cite{devlin2018bert} was pre-trained on large unlabeled text corpora,  
before being fine-tuned and achieving state-of-the-art performance on eleven downstream tasks (e.g., classification, similarity and paraphrase, inference).
Variants of BERT have since been developed, and outperformed state-of-the-art solutions~\cite{yang2019xlnet, liu2019roberta, sanh2019distilbert}. 

In addition to the substantial performance increases, foundational models significantly reduce and even eliminate the need for data labeling, a process often considered tedious, error-prone, and expensive. The pre-training phase which is the most compute intensive phase is performed on unlabeled data in a self-supervised manner, and only the fine-tuning phase requires a small amount of labeled instances. Intuitively, in the pre-training phase, the models extract general useful features from raw text because of the large volume of unlabeled data that is usually available much more readily than labeled data (e.g., the common crawl corpus has billions of tokens and covers multiple languages). In this pre-training phase, the model learns numerical representations for words that capture semantic information, and relational knowledge behind them. For example, words belonging to a same lexical field, such as \textit{Human, Face, Speech, Body}, have representations that tend to be closer~\cite{turton-etal-2021-deriving}. In addition, foundational models learn to discern the different meanings a same word can have based on its context. For example, they will generate different numerical representations for the same word \textit{die} in the following two sentences: ``Many more \textit{die} from radiation sickness, starvation and cold'', and ``Players must always move a token according to the \textit{die} value''~\cite{NEURIPS2019_159c1ffe}. Then, after having learnt general linguistic patterns, the models can be fine-tuned on specific downstream tasks, with considerably fewer labeled instances.
GPT-3~\cite{brown2020language}, a more recent language model, even eliminates the need for fine-tuning; and reduces the amount of labeled examples further by one order or magnitude, and even eliminates the need for any labeled data for some tasks.
In addition to NLP, foundational models are being applied in vision~\cite{dosovitskiy2020image, pmlr-v119-chen20s, radford2021learning}, tabular data~\cite{yin2020tabert}, and  programming languages~\cite{pmlr-v119-kanade20a, feng-etal-2020-codebert}.


We observe that network traffic analysis, and management share a number of properties with NLP and other domains where foundational models have been very successful:

    \noindent $\bullet$ {\em Similarities of downstream tasks:} First, similar to the status in NLP prior to the emergence of foundational models, specific solutions with their own preprocessing, features extraction, architectures, and datasets, are currently being developed for different tasks 
    (e.g., congestion control~\cite{10.1145/2486001.2486020,  10.1145/3387514.3405892, pmlr-v97-jay19a}, adaptive bitrate streaming~\cite{10.1145/3098822.3098843}, datacenter-scale automatic traffic optimization~\cite{10.1145/3230543.3230551}, job scheduling~\cite{krishnan2018learning, 10.1145/3341302.3342080, 10.1145/3341302.3342080}, resource management~\cite{10.1145/3314148.3314357, 9068634, conf/icnp/ZhangBLMLHX19}, network planning~\cite{10.1145/3452296.3472902}, packet classification~\cite{10.1145/3341302.3342221}, performance prediction~\cite{10.1145/3387514.3405868}, congestion prediction~\cite{ DBLP:conf/icnp/PoularakisQLKT21}, performance estimation~\cite{10.1145/3452296.3472926}, malware detection~\cite{kharraz2018surveylance, xu2021detecting}, mapping from a low-quality video to a high-quality version \cite{10.5555/3291168.3291216}, or semi-automated generation of protocol implementations from
    specification text~\cite{10.1145/3452296.3472910}.)
    Next, we observe that most of the underlying adopted machine learning approach  behind those solutions (e.g., classification, anomaly detection, generator, and reinforcement learning) are areas where foundational models have been successfully applied, or being explored (Section~\ref{sec:tasks}).

    \noindent $\bullet$ {\em Abundance of unlabeled data:} Second, there is abundant unlabeled network data: For example, universities' networks have reported 1 to 2 TB of traffic on average per day~\cite{5375520, 10.5555/1251086.1251109}; and data-center networks and content providers generate more than 10 TB on average daily~\cite{10.1145/1644893.1644918, 10.1145/3487552.3487860, 10.1145/2068816.2068837} (Section~\ref{sec:data}).
    
    \noindent $\bullet$  {\em Rich semantic content:} Third, network data has rich semantic information: For example, intuitively, for transport protocols, SCTP, TCP, and MPTCP may be considered more similar than UDP, since they handle congestion and packet loss recovery; for applications, HTTP and CoAP may be closer to each other than SMTP, IMAP, or POP3; for security, ciphersuites may form clusters (e.g., weak versus strong). In NLP, foundational models have been very effective in capturing semantic relationships behind, and between words~\cite{conf/aaai/BouraouiCS20} (Section~\ref{sec:semantics}).

We highlight early efforts that begun exploring foundational models in the networking domain, and which show promising results (Section~\ref{sec:success}).
However, to  fully realize the benefits of foundational models, we identify a  number of challenges network data present, and that must be addressed:. 

     
     
    \noindent $\bullet$   {\em Extraction of general useful patterns from networking data:} Although network data has rich semantic content, how can we effectively extract it? 
    First, we observe there are few dozens popular deployed network protocols (e.g., TCP, UDP, HTTP, SMTP, NTP, SIP, RTSP, DHCP, DNS, etc.) which have rich semantic relationships between and across them: For example, TCP and UDP are transport protocols, and HTTP runs on top of TCP. As such, can we define a single cross-protocol representation that capture those semantic relationships, and that can significantly improve the performance for a wide range of networking tasks (Section~\ref{sec:representation})?
    Next, \textit{tokenization} (Section~\ref{sec:tokenizer}), \textit{context} (Section~\ref{sec:context}), and \textit{pre-training tasks} (Section~\ref{sec:pre-training}) have played critical roles towards effectively extracting semantic relationships in NLP. 
    What would their counterparts for networking data be, especially given its unique characteristics: e.g., packets from different connections may be interleaved, or fields (e.g., HTTP User-Agent, DNS answers) may contain different data structures (e.g., text, sets)?
      
    \noindent $\bullet$   {\em Publicly available datasets:}  The public availability of large text corpus (e.g.,~\cite{zhu2015aligning, raffel2020exploring}, and labeled benchmarks (e.g., \cite{wang2018glue, rajpurkar-etal-2016-squad, zellers2018swag}) with a wide variety of downstream tasks, and groundtruth has been essential for research, and making advances in the emergence of foundational models and NLP. Although organizations can collect and analyze large amount of their own network data, can the networking community release training data available, and define networking benchmarks (Section~\ref{sec:benchmarks})?
      
    \noindent $\bullet$   {\em Dealing with rare and unseen events:} Reports raised uncertainty about the suitability of machine learning for network security, and more specifically anomaly detection~\cite{5504793}, claiming that
    the task of detecting attacks may be fundamentally different from
    other applications where machine learning have been successfully deployed. Can foundational models therefore effectively help detect zero-day attacks as well as unusual behaviors (Section~\ref{sec:security})?

    \noindent $\bullet$   {\em Interpretability:} 
    Understanding the reasons behind a model's prediction outcome allows users to validate models, and increases users' trust in the models~\cite{lapuschkin2019unmasking, 10.1145/3232565.3232569}.  
    Researchers have argued for interpretation methods specifically developed for networking models~\cite{meng2020interpreting} given the unique nature of networking inputs. Can we derive meaningful explanations for foundational models when applied to networking (Section~\ref{sec:interpretability})?


  We elaborate on these challenges and identify  opportunities to tackle them. 
  Our  paper is  a ``call to arms''  for exploring the potential benefits of foundational models to the domain of networking. As such, our goal here  is to ignite  this discussion, and we acknowledge that we raise more questions than provide concrete     answers.
  

\section{Background}
\label{sec:background}

The goal of this section is to given an overview on  foundational models. More specifically, we describe how BERT models are trained, and how they learn numerical representations (i.e., high-dimensional vectors) -- also called \textit{embeddings} -- for words. However, before going into the details of BERT models, we first briefly discuss why embeddings were introduced, and the solutions that were first developed. 

\textbf{Motivation:} Neural networks cannot operate on text (e.g., string) directly. They require all inputs to be numerical. As such, text has to be converted into numerical values. One hot encoding, a popular method for categorical variables, assigns each value (e.g., ``conference'', ``workshop'', ``journal'') to a binary vector (e.g., 00...0001, 00...0010, 00...0100). As such, every pair of values is equidistant. Instead, for text, words with similar meanings should have similar vectors.

\textbf{Word2Vec:} To satisfy this requirement, Word2Vec was introduced in 2013~\cite{mikolov2013efficient}.  Relying on two neural network variants, it computes
word embeddings based on the words’ context: Continuous
Bag-of-Words (CBOW) predicts the current word based on
the context; and Skip-gram instead predicts the closely related
context words to an input word. Word2Vec can as such learn high dimensional (e.g., 50, 100, 300, 600) vector representation for words, and was shown
to learn
very subtle semantic relationships between words, such as
a currency and the name of the country that uses it, e.g. Angola is to kwanza as Iran is to rial~\cite{mikolov2013efficient}, or that 
``King - Man + Woman'' resulted in a vector very close
to ``Queen''~\cite{mikolov-etal-2013-linguistic}.

\textbf{BERT:} 
In 2018, Devlin \textit{et al.} presented BERT 
(Bidirectional Encoder Representations from Transformers), a new language representation model~\cite{devlin2018bert}.
In contrast to Word2Vec which computes context-independent embeddings, BERT generates contextual embeddings. To illustrate the differences, we consider the two following sentences: ``\textit{Bark} is essential for a tree's survival'', and ``There can be many reasons behind a dog's \textit{bark}''. Word2Vec would generate the same vector representation for the word \textit{bark} -- independently of its position and meaning -- in both sentences. In contrast, BERT would generate different vector representations for each occurrence of \textit{bark} because their position, and surrounding words differ.

More generally, training BERT relies on two stages: pre-training and fine-tuning:

\noindent $\bullet$ \textit{Pre-training}: The model is trained on unlabeled data over two pre-training tasks. In the first task, called, Masked Language Modeling, a fraction of the input
tokens are randomly masked, and the goal is to predict those masked tokens. In the second task, Next Sentence Prediction, two sentences, \texttt{A} and \texttt{B}, are provided as inputs to the model, and the goal is to predict whether \texttt{B} is the actual sentence that follows \texttt{A} in the initial corpus, or a random sentence from the corpus.

\noindent $\bullet$ \textit{Fine-tuning}: To fine-tune a BERT model to a downstream task, an additional layer is typically added to the pre-trained model, and then the entire model is trained over the labeled data for few epochs. 

\textbf{GPT-3:} Generative Pre-trained Transformer 3 (GPT-3) is Transformer-based language model with 175 billion
parameters, developed by OpenAI in 2020~\cite{brown2020language}. It further reduces the amount of required labeled data during the fine-tuning stage, from thousands or tens of thousands of examples, to tens or hundreds of examples (few shot learning), one single example (one shot learning), or no example and only an instruction in natural language is given to the model (zero shot learning). Contrary to the fine tuning stage, GPT-3 does not perform any gradient update. Instead, the model applies ``in-context learning'', where the model is simply conditioned on a natural language instruction and/or a few examples of the task. 
More specifically, as input (also, called the prompt), the model is provided the instructions and/or labeled examples. In return, the model generates a text completion: As illustrated in the Open AI API documentation, given the prompt, ``Write a tagline for an ice cream shop'', the model returns the following completion: ``We serve up smiles with every scoop!''
In NLP tasks, GPT-3 demonstrated promising results in zero-shot and one-shot settings; and in few-shot setting, it was sometimes competitive with or even occasionally outperformed state-of-the-art models.

\section{A Case for  Network  Foundational Models}

We point out similarities between networking, and other domains (e.g., NLP) where foundational models had a large impact; and recently published early work that provide corroborating evidence of the potential of foundational models for the networking domain.

\subsection{ Range of  downstream tasks}
\label{sec:tasks}

A wide range of network downstream tasks can benefit from foundational models. First, we observe that  machine learning solutions have been developed for different network downstream tasks, including towards congestion control~\cite{10.1145/2486001.2486020,  10.1145/3387514.3405892, pmlr-v97-jay19a}, adaptive bitrate streaming~\cite{10.1145/3098822.3098843}, datacenter-scale automatic traffic optimization~\cite{10.1145/3230543.3230551}, job scheduling~\cite{krishnan2018learning, 10.1145/3341302.3342080, 10.1145/3341302.3342080}, resource management~\cite{10.1145/3314148.3314357, 9068634, conf/icnp/ZhangBLMLHX19}, network planning~\cite{10.1145/3452296.3472902}, packet classification~\cite{10.1145/3341302.3342221}, performance prediction~\cite{10.1145/3387514.3405868}, congestion prediction~\cite{ DBLP:conf/icnp/PoularakisQLKT21}, performance estimation~\cite{10.1145/3452296.3472926}, malware detection~\cite{kharraz2018surveylance, xu2021detecting}
and , semi-automated generation of protocol implementations from
specification text~\cite{10.1145/3452296.3472910}.

Next, we note that those solutions can be classified by the underlying adopted machine learning approach such as classification, anomaly detection, generator, and reinforcement learning; and most are areas where foundational models have been very successfully applied, or where foundational models are currently being expanded to: For example, foundational models have set state-of-the-art performance for text classification~\cite{liu2019roberta}, token classification~\cite{sanh2019distilbert}, and text generation~\cite{brown2020language}; and foundational models have recently also been applied to reinforcement learning problems~\cite{chen2021decision, janner2021offline}, or translation from natural language to code~\cite{chen2021evaluating}.  
Codex~\cite{chen2021evaluating} is a GPT-3 language model fine-tuned on publicly available code from GitHub. Given a coding task in natural language as a prompt, the model returns blocks of code that satisfies it. This task can be considered to be similar to the network goal of generating protocol implementations from specification text~\cite{10.1145/3452296.3472910}.

\subsection{Abundant unlabeled data}
\label{sec:data}

Unlabeled data is plentiful in network: For example, universities' networks carry 1 to 2 TB of traffic on average per day~\cite{5375520, 10.5555/1251086.1251109}; and data-center networks and content providers generate more than 10 TB of traffic on average daily~\cite{10.1145/1644893.1644918, 10.1145/3487552.3487860, 10.1145/2068816.2068837}.
In comparison, 
BERT was trained on 16 GB of Books Corpus and English Wikipedia~\cite{devlin2018bert}, RoBERTa uses 160 GB of text for pre-training~\cite{liu2019roberta},  XLM-RoBERTa was trained from 2.5 TB of text~\cite{conneau-etal-2020-unsupervised}, and the OpenAI GPT-3 model was trained on 45TB of text~\cite{brown2020language}.

\subsection{Rich semantic content}
\label{sec:semantics}

We envision foundational models to be applied to network data, and we argue that similar to text, network data has rich semantic content that the pre-training phase would be able to extract, and make it useful to a wide range of downstream tasks.
More specifically, a packet trace can be viewed as a sequence of variables, some of which can be categorical, and others numerical. For example, the \textit{total length} field in an IP header is a 16-bit numerical variable that indicates the entire size of the IP packet (header and data) in bytes. 
In contrast, 
the \textit{protocol field} in the IP header is an 8-bit categorical variable that indicates the next protocol inside the IP packet. Possible values include TCP, UDP, ICMP, SCTP, EIGRP, DSR, IPv4, IPv6, and GRE; and one can observe that they can form semantic clusters, e.g., with TCP, UDP and SCTP being transport protocols, EIGRP and DSR being routing protocols, and IPv4, IPv6, and GRE indicating tunneling. Another example of categorical variable with rich semantic information is the DNS query field. Values may indicate mail servers (e.g., gmail.com, outlook.com), repository servers, time servers (e.g., time.nist.gov, ntp.org), news sites (e.g., npr.com, nytimes.com), or video streaming sites (e.g., netflix.com, primevideo.com).





\subsection{Early successes}
\label{sec:success}

We summarize  results from some early work exploring  the potential of foundational models for the networking domain.




\textbf{NetBERT:}
A recent study~\cite{louis2020netbert} provides evidence that network data includes rich semantic relationships. The authors trained a BERT model on a  text corpus on computer networking. They do not apply foundational models directly to network data. However, the results reveals subtle relationships in the networking domain.
For example, similar to  ``Man is to King as Woman is to Queen''  (Section~\ref{sec:background}), the authors confirmed several similar analogies in the network domain including ``BGP is to router as STP is to switch'', ``MAC is to switch as IP is to router'', or ``IP is to network as TCP is to transport''.


\textbf{NorBERT:} A more recent study explored the adaptation of foundational models on network data, and preliminary results provide additional evidence of the semantic richness of network data, and performance improvement for networking downstream tasks~\cite{norbert}.

    \noindent $\bullet$ {\em Semantic relationships in networking data:} Adapting foundational models on networking data, the authors revealed  interesting relationships between tokens' embeddings. For example, the closest neighbor to the token $80$ (HTTP), was the token $443$ (HTTPS); and the closest neighbor to the token $49199$  (ciphersuite ``ECDHE + RSA authentication AES-128 GCM SHA-256''), is token $49200$ (ciphersuite ``ECDHE + RSA authentication AES-256 GCM SHA-384''). These two ciphersuites  differ only in the keys' lengths. The closeness of these tokens is according to intuition, and domain knowledge. 
    
    \noindent $\bullet$ {\em Performance improvement:} The authors also compared the performance of foundational models for  downstream classification tasks. The authors pre-trained a foundational model (NorBERT) on DNS traffic, fine-tuned it on a labeled dataset, and  evaluated its performance on an independent labeled dataset. The performance are compared with those of gated recurrent units (GRU) models, with both initialization to random values, and context-independent embeddings (GloVe)~\cite{pennington2014glove}.
The performance of the GRU models drop considerably on the validation dataset (F-1 scores between 0.585 and 0.726). In contrast, the performance of NorBERT remains above 0.9, demonstrating significant performance improvements.



\section{Challenges and Opportunities}


Despite the promising results of early efforts, 
we identify broader challenges that network data presents given their unique characteristics. More specifically, NetBERT essentially provided evidence that the networking domain contains rich semanctic relationships, but the study applied foundational models on text related to networking, and not networking data; and NorBERT demonstrated significant performance gains, but focused on two network downstream classification tasks. For foundational models to be applicable, and useful to a broader range of networking downstream tasks, 
we identify challenges, 
and opportunities to tackle them.

\subsection{Extraction of general useful patterns}

\subsubsection{Common Representation}
\label{sec:representation}


Akin to languages, there are a hundred popularly used network protocols. A network protocol is a language of communication between two entities (e.g., client and server). An intrinsic property of a language is that an utterance between two entities at a certain point in space and time, bears a semblance to utterances between two other entities at an entirely different points in space and time. Consider a client and a web server communicating via the HTTP protocol; interactions between any pair of client and web server using the HTTP protocol (language) bears similarities between each other. For example, given a certain utterance from the client (e.g., HTTP GET), there are certain sets of valid utterances (responses) from the web server (e.g., STATUS 200). A wider context such as knowing the HTTP User Agent type or the size of the HTTP response, helps predict future utterances by both the client and server. 

One key distinction between network protocols and natural languages is that network protocols are almost always multiparty (at least two) communication. The topology between the parties involved is an important aspect of the network protocol. Many of the foundation models in NLP are focused on monologues (e.g., document) or two party communication (e.g., chat bot). However, it is very common for an application to span multiple network protocols, and to involved multiple servers, to achieve a single task. Hence, a natural first step is for us to learn common representations within a single network protocol and then expand the foundation model to the multi-lingual domain that captures multi-party, multi-protocol applications (akin to the evolution in NLP domain from Roberta~\cite{liu2019roberta} to XLM-Roberta~\cite{conneau-etal-2020-unsupervised}, a transformer-based multilingual
masked language model pre-trained on text in 100
languages.)

\subsubsection{Tokenizer}
\label{sec:tokenizer}


Tokenization is an important step in text pre-processing. It splits a piece of text (e.g., phrase, sentence, paragraph) into smaller units, called tokens. 
Tokens represent the smallest semantic unit of interest.
A token can be a word, a subword, or even a character. 
Foundational models frequently use subword-based tokenization algorithms. More specifically, BERT~\cite{devlin2018bert} uses WordPiece~\cite{6289079}, and RoBERTa~\cite{liu2019roberta} uses Byte-Pair Encoding~\cite{sennrich-etal-2016-neural}.
Text is split into words based on delimiters, with common ones being space and punctuation; and to reduce the vocabulary size, to handle rare words, and to learn meaningful representations, words may be split into sub-words. For example WordPiece would split the word `\texttt{Hotnets}'' into two tokens [`\texttt{hot}', `\texttt{\#\#nets}'].
Studies have revealed that 
subword tokenization plays an important role in the relationships that models learn from corpus~\cite{singh-etal-2019-bert}. 

However, with packet traces being often viewed as sequences of bytes, with no clear delimiters such as white spaces, and punctuation, how should network data therefore get tokenized?
One approach could consist in applying character-based tokenizers~\cite{kalchbrenner2016neural, lee2017fully, radford2017learning}. Another approach may consist in recognizing the network protocol (language) and tokenizing it based on protocol format (e.g., 4 byte IP address, 2 byte port number, one byte TCP flag, HTTP fields, etc.). This would preserve the semantics of the tokens as per the underlying network protocol specifications. 


\subsubsection{Context}
\label{sec:context}


Word embeddings are based on the premise that words that occur in similar contexts tend to have similar meanings. For example, Word2Vec trains two neural networks to predict a word given its context and vice-versa~\cite{mikolov2013efficient}; and foundational models are generally trained using a masking approach: The key idea is that given a sequence of tokens (often called the context), a random percentage (typically 15\%) of the tokens are masked and the model is trained to reconstruct the masked tokens. A byproduct of this training process is embeddings, which find application in various downstream tasks. This introduces the challenge of defining a context around tokens. 

First, it may be possible to define a context based on packet boundaries (shorter context) or session boundaries (wider context).
In addition, we observe that at a point of packet capture (e.g., border router), packets from different end points may be interleaved. Even when focusing on traffic from and to individual end points, their traffic may consist of packets belonging to concurrent connections. Focusing on individual connections may lose semantic relationships between connections, especially given that a transaction may consist of concurrent and sequential connections. 
Third, practical constraints may limit the size of context to about 512 tokens. 
As such, it may be vital to construct non-standard contexts over network protocols: e.g., use the first $M$ tokens from each of the $N$ successive IP packets sent or received from an endpoint as a context.

\subsubsection{Pre-training tasks}
\label{sec:pre-training}

What pre-training tasks would be most effective for network data?
BERT defined two pre-training tasks: Masked language modeling, and next sentence prediction (Section~\ref{sec:background}). 
%
Since then, a number of studies explored alternative pre-training tasks~\cite{zhang2019ernie, ling-etal-2015-design, choi-etal-2018-ultra, han-etal-2018-fewrel, zhang-etal-2017-position, sun2020ernie, wang2018glue, joshi2020spanbert, pradhan-etal-2012-conll, wang2019structbert, feng-etal-2020-codebert, Clark2020ELECTRA:}.

Given the network-specific downstream tasks (Section~\ref{sec:tasks}), and characteristics of network data (Sections~\ref{sec:representation}, and~\ref{sec:context}), new network-specific training tasks may need to be defined. For example, query-answers are common transactions in computer networks, and new training tasks may be required to capture the nature of the relationships between a query and its answers: In DNS, the answers may be viewed as the children of the query in a hierarchical tree. In addition, network fields may consist of different structures, such as sets where the order does not matter. An example is the DNS answer field, where multiple values may be returned for a single query. Such structures (e.g., list, set) may reflect stronger similarities between their members.

\subsection{Publicly available data and benchmarks}
\label{sec:benchmarks}

The public availability of large text corpus (e.g.,~\cite{zhu2015aligning, raffel2020exploring}, and labeled benchmarks (e.g., \cite{wang2018glue, rajpurkar-etal-2016-squad, zellers2018swag}) 
are one of the main reasons behind the active and prolific research in foundational models and NLP.
For example, the General Language Understanding Evaluation (GLUE) benchmark~\cite{wang2018glue} includes a collection of natural language understanding tasks including classification tasks (e.g., is a sentence grammatically acceptable?~\cite{warstadt2019neural}, or is a review positive?~\cite{socher-etal-2013-recursive}), similarity and paraphrase tasks~\cite{dolan-brockett-2005-automatically, cer-etal-2017-semeval} (e.g., are two questions semantically equivalent?), and inference tasks~\cite{rajpurkar-etal-2016-squad, levesque_winograd_2012} (e.g., does a premise sentence entail or contradict a hypothesis sentence?~\cite{williams-etal-2018-broad}). 

Although network data is abundant (Section~\ref{sec:data}), and organizations can easily collect their own, concerns of leaking sensitive content have limited their public release. To address the lack of public networking data, researchers set up small private labs with a variety of devices, collect their traffic, and publicly release the packet captures (e.g.,~\cite{unsw}). However, can the networking community offer larger amount of network data, and define benchmarks to facilitate research? Synthetic packet traces generators~\cite{10.5555/1762888.1762896, 10.1145/1005686.1005733, 10.1145/1151659.1159928, 10.1145/1140086.1140094, xu2021stan} may be one solution for mitigating the privacy concerns, and training foundational models on network data.
Benchmarks could comprise a dozen of network downstream tasks including device classification, flow classification, performance prediction, congestion prediction, malware detection.

\subsection{Rare and unseen events}
\label{sec:security}


Sommer and Paxson~\cite{5504793} reported that contrary to several other domains where machine learning was successfully commercially deployed (e.g., recommendation systems~\cite{1167344, bennett2007netflix}, optical character recognition systems~\cite{10.5555/1304596.1304846}), machine learning was rarely deployed in operational network settings for anomaly detection. The authors argue that that ``\textit{the strength of machine-learning tools is finding activity that is similar to something previously seen}'', and not to find ``\textit{novel attacks}''.

However, methods have recently been developed specifically to detect out-of-distribution instances, i.e., instances that differ from those seen during training (e.g.,~\cite{10.5555/3045390.3045502, 10.5555/3295222.3295387, liang2017enhancing, 10.5555/3327757.3327819, liang2017enhancing, 10.5555/3327757.3327819, liu2020simple, 10.5555/3495724.3497526}), and machine learning algorithms are now commercially deployed to identify defects (e.g., in car assembly line) where the costs of errors are also high. In other words, while machine learning may not have been suitable for network intrusion detection, recent advances in out-of-distribution may help effectively identify zero-day attacks, and unusual network behaviors.

\subsection{Interpretability}
\label{sec:interpretability}

Interpretability is critical towards validating models, and increasing users' trust in the models~\cite{lapuschkin2019unmasking, 10.1145/3232565.3232569}. As such, a large amount of research effort has been devoted towards understanding the reasons behind a model's prediction outcome (e.g.,~\cite{sundararajan2017axiomatic, smilkov2017smoothgrad, lapuschkin2019unmasking, lundberg2017unified, ribeiro2016should, serrano2019attention, jain2019attention, wiegreffe2019attention, 9671639}). 
However, machine learning solutions for networking downstream tasks often work with domain specific inputs. For example, solutions may take a topology as input, and return routing paths~\cite{rusek2019unveiling}. Therefore, researchers have argued for interpretation methods specifically developed for networking models~\cite{meng2020interpreting}. Similarly, we argue that interpretability methods specifically developed for foundational models applied to networking may be required. For example, if networking data is tokenized at the character level (Section~\ref{sec:tokenizer}), how can we derive meaningful explanations? To draw an analogy with computer vision, the notion of superpixels (i.e., set of adjacent pixels with similar color and underlying properties)~\cite{STUTZ20181} has allowed more meaningful features and explanations~\cite{ribeiro2016should, ghorbani2019towards}.


\subsection{Other Issues}
\label{sec:roadmap}


The above challenges explore  only a small sample of open issues. We conclude by identifying a few other open issues:
\begin{packeditemize}
\item {\em Energy footprint:} Large models training and inference often consume massive amount of energy, raising questions on the benefits and generability of those models. 
\item {\em Generalizability:} Although many downstream tasks can potentially benefit from common representations, we do not expect a single universal foundation model to be able to address most relevant tasks. Instead, distinct models may be required for different areas (e.g., security, resource management). As such, what is the minimum coverage set of foundation models that can cover the relevant desired tasks? 
\item {\em Learning complexity:}  The dimensionality of  networking data could be larger than that of text? If so, what would the required dimension of the embeddings be; and how much training data would be required?
\end{packeditemize}

\section*{Acknowledgments}
We thank the reviewers for their insightful feedback.
\bibliographystyle{ACM-Reference-Format} 
\bibliography{hotnets22}

\end{document}